\documentclass{webofc}
\usepackage[varg]{txfonts}   % Web of Conferences font
\usepackage{graphicx}
\usepackage{hyperref}

\begin{document}

\title{New ADS Functionality for the Curator}

\author{\firstname{Alberto} \lastname{Accomazzi}\inst{1} \fnsep\thanks{\email{aaccomazzi@cfa.harvard.edu}} \and
        \firstname{Michael J.} \lastname{Kurtz}\inst{1} \and
        \firstname{Edwin A.} \lastname{Henneken}\inst{1}
        \and
        \firstname{Carolyn S.} \lastname{Grant}\inst{1} \and
        \firstname{Donna M.} \lastname{Thompson}\inst{1} \and
        \firstname{Roman} \lastname{Chyla}\inst{1} \and
        \firstname{Steven} \lastname{McDonald}\inst{1} \and
        \firstname{Taylor J.} \lastname{Shaulis}\inst{1} \and
        \firstname{Sergi} \lastname{Blanco-Cuaresma}\inst{1} \and
        \firstname{Golnaz} \lastname{Shapurian}\inst{1} \and
        \firstname{Timothy W.} \lastname{Hostetler}\inst{1} \and
        \firstname{Matthew R.} \lastname{Templeton}\inst{1}
}

\institute{Harvard-Smithsonian Center for Astrophysics, 60 Garden St., Cambridge, MA, USA.}

\abstract{%
In this paper we provide an update concerning the operations of the NASA Astrophysics Data System (ADS), its services and user interface, and the content currently indexed in its database.  As the primary information system used by researchers in Astronomy, the ADS aims to provide a comprehensive index of all scholarly resources appearing in the literature.  With the current effort in our community to support data and software citations, we discuss what steps the ADS is taking to provide the needed infrastructure in collaboration with publishers and data providers. A new API provides access to the ADS search interface, metrics, and libraries allowing users to programmatically automate discovery and curation tasks.  The new ADS interface supports a greater integration of content and services with a variety of partners, including ORCID claiming, indexing of SIMBAD objects, and article graphics from a variety of publishers.  Finally, we highlight how librarians can facilitate the ingest of gray literature that they curate into our system. 
}
\maketitle
\section{Introduction}
\label{intro}
The NASA Astrophysics Data System plays a central role in the ecosystem of information providers in Astronomy.  As the main discovery platform for the scientific literature used by the community of research astronomers, ADS attempts to provide a timely and complete record of scholarly artifacts for our discipline.
One of the features that makes ADS such a compelling platform for a scientist is not just its disciplinary completeness but also the richness of the information that it collects.  Rather than simply serving as a library catalog, ADS enriches its bibliographic records via the integration of content from a variety of data providers, including links to datasets hosted by NASA and other archives and bibliographies maintained by collaborating curators and librarians.

As an information system, ADS has become a platform where decisions about the inclusion and classification of both conventional and non-traditional content has become a community-driven effort.  The integration of community-curated content and the need to support the needs of librarians which maintain such content is so important that both these activities are explicitly mentioned in ADS's official mission statement.
Our participation in conferences such as LISA and in the conversations surrounding these activities is a testament to the importance of the partnership between ADS and the library community.

\section{Content}
\label{content}
ADS aggregates content from a variety of scholarly information providers, and adds value to such content by enriching and linking its records with a wealth of internal and external resources.  
The policies and decisions which inform what content should be indexed in the system are generally dictated by the needs of the research community in Astronomy.  As these needs and expectations change, so do our policies.

An example of such change is the way in which arXiv eprints have been ingested and used in ADS.  Originally these publications were indexed as a separate collection of records available for search on demand, and lived alongside the traditional literature indexed in ADS.  At a later time, ADS started integrating references published in eprints in its citation database, so that citations from eprints would contribute to the total (non-refereed) citations accrued by published papers.  Finally, once we started cross-identifying eprints with the corresponding published records, citations made by and to eprints were automatically attributed to the corresponding published papers.  Today, ADS's new search engine and user interface simply consider eprints as an additional source of publications which may or may not overlap with the refereed literature.

Similarly, we are now seeing an evolution in the way a variety of non-traditional content is considered not only relevant but rather essential to the research process.  Transparency, reproducibility, and credit are the three major reasons behind current publishing trends which encourage data and software citation \cite{scwg, dcwg}.  While not traditional articles, all of these resources have begun appearing in citation lists of papers published in Astronomy, and as our community is moving to support data and software citations \cite{2016arXiv161106232T}, it becomes crucial for ADS to support these efforts.  By making sure that whatever citable scholarly astronomy resource is indexed in our system, ADS makes these resources easily discoverable, while at the same time giving them the proper citation credit.  Such content expansion does not come without additional curation challenges, imposing new requirement on the workflows managed by the project.  We are currently working with the AAS and Zenodo on a Sloan-funded effort to implement software citation indexing in ADS.

We are also seeing a rapid change in the way individuals and organizations "publish" content.  As more options become available to scientists and librarians alike to make a variety of scholarly content available on a variety of digital platforms, the question remains of how discoverable and citable this information may be.  As our community looks to ADS as the primary search engine for scholarly astronomy content, it becomes essential that new scholarly publication venues be included in what we index.  We are currently seeing the rise of general-purpose repositories such as \href{https://zenodo.org}{Zenodo} as publishing platforms for gray literature, software, and data products.  We have started ingesting such material to the extent that we can determine its relevancy and reputation.  Our partnership with librarians in vetting specific communities (such the Zenodo Astronomy Thesis Collection) and creating specific event venues (such as presentations and posters given at a particular conference) provides the appropriate level of trust in this material which would be otherwise onerous to evaluate.  If you are supporting the publication and archiving of such content please contact us to ensure that it gets properly ingested in ADS.

Finally, we continue to collaborate with a number of different archives and institutions in the maintenance of bibliographic groups which represent collections of papers belonging to either an institutional or so-called "telescope" bibliographies \cite{2012SPIE.8448E..0KA}.  Such collections, once indexed in ADS, provide a very effective and powerful way to select papers containing observations from a particular telescope or mission (e.g. HST), or published by scientists at a particular institution (e.g. CfA).  Their existence makes it easy to generate useful statistics about missions, projects and research centers, understand the science being enabled or pursued by them, and even locate the datasets which were generated or analyzed by them.

\section{Services}
\label{services}

The ADS is built around a database, and access to its contents is available through a number of services.  These services are  used by a number of different applications, some of which act on behalf of individual users (e.g. somebody interacting with the system through a web browser), others on behalf of curators (e.g. a librarian using a tool to maintain a bibliography), scholarly information providers (e.g. a publisher validating bibliographic links), or even search engines (e.g. the Googlebot crawler).  The technical challenge in maintaining these services is providing the proper level of responsiveness and functionality to all our constituents.  In this respect, prioritization of tasks and features is not always a simple matter: while a bug or the lack of a feature may only affect a few librarians, its impact may be great overall, if it were to affect the curation of an entire data center bibliography or the access to its data holdings. 

Managing a system such as the ADS also presents administrative and legal challenges such as protecting the privacy of its users and safeguarding its content, while at the same time providing useful services to the community.  Since data is given to the ADS under specific terms of usage from each publisher, we need to make sure that our search services and interfaces protect this content from abusive behaviour such as the indiscriminate download of most of our records or full-text papers by any one individual.  From a technical prospective, this has led us to the implementation of rate limits for all of our new services: each (possibly anonymous) user is allowed only N amount of queries for any given period of time. There is often a fine line between securing the system against misuse and providing users with the access they need to conduct research studies based on ADS content.  In general, we aim to have N be large enough that no individual researcher interacting with our system on a daily basis will ever run out of available queries, but that applications which systematically query and download records from ADS might.  If you find yourself in need to have these limits raised for legitimate reasons, feel free to contact us with details of what you are trying to accomplish and we will work with you to make sure you can get your work done.

\subsection{Application Programming Interface}
\label{api}
Access to the ADS services is delivered through a so-called RESTful Application Programming Interface (API), which provides \href{https://api.adsabs.harvard.edu}{a single, top-level entry point} to all of its services.
The API supports self-discovery of all its services and corresponding endpoints via a \href{https://api.adsabs.harvard.edu/resources}{versioned public URL}.
Internally, the API connects to a number of different microservices which implement parts of the ADS system.  Among them one can find interfaces to search, metrics, user libraries, ORCID, visualizations, export, and recommender services.  We expect this list to grow over time.

Access to the API and the underlying services is best accomplished through libraries and applications.  For those curators interested in programmatic access to the API, \href{https://github.com/adsabs/adsabs-dev-api}{we provide online documentation} and recommend the use of existing libraries such as \href{http://ads.readthedocs.io/en/latest/}{the ads python module}, which is available for direct install from PyPI.

\subsection{Search Engine}
\label{search}
At the core of ADS's system is a search engine which provides access to the entirety of the corpus indexed by it.  The current search engine is an ADS-modified version of the open-source project Apache SOLR, which is an enterprise level search solution used by a large number of academic institutions and commercial companies.  The adoption of a robust, open source system such as SOLR has allowed us to index for the first time not only the main article metadata, but also the full-text content of the articles in our possession.  This has increased the amount of data indexed in our system by one order of magnitude compared to the prior generation search.

While most users limit their searches to the basic article metadata (author, title, keywords and abstract), the implications of being able to perform full-text searches is a game-changer for curators who have a need to find mentions of facilities, telescopes, instruments or even specific observations in the literature.  ADS-specific enhancements to SOLR give users easy access to the citation and usage data indexed in the system via so-called second-order operators.  As an example, the \emph{citations(query)} operator allows retrieval of all articles citing the results returned by the inner search \emph{query}.  Additional operators allow retrieval of trending, useful, and review articles on a particular topic.

\subsection{User Interface}
\label{ui}
In 2015, the ADS project launched a new User Interface, which was code-named ADS Bumblebee.  Bumblebee represents the latest product in the evolution of our UI and attempts to integrate the best practices from the fields of usability and graphic design.  Our goal was to create an interface that is intuitive and forgiving while simultaneously providing access to the full range of ADS's sophisticated query and analytical capabilities.  

Bumblebee is entirely a javascript application and as such runs on a user's web browser, requesting content from the ADS services via the API described above.  In that sense, it is nothing more than an application built on top of the public ADS infrastructure, and as such could be modified and improved by anybody. The interface integrates and highlights a number of useful features offered by the API services, among which are the availability of facets (filters) for query refinement, scalable metrics for any set of query results, and detailed metrics for collections of records.
A number of visualizations have been implemented that provide additional insights into the search results.  Visualizations of author networks and co-citation networks provide a hierarchical view of research groups and research topics, respectively. Displays of paper analytics provide plots of the basic article metrics (citations, reads, and age).  All visualizations are interactive and provide ways to further refine search results.
The detailed view of an individual article now includes a preview of the images and plots included in the paper thanks to our collaboration with publishers such as the AAS, EDP Sciences, and Elsevier.  

Of particular interest to librarians and curators is the inclusion in any search results of features provided by the SOLR search engine and our new API.  These include the highlighting of search terms in context, a histogram of the publication dates for any set of results, the distribution of citation counts, and a breakdown of records by collection, bibliographic group, or keywords.

\subsection{ORCID}
\label{orcid}
The new ADS interface and search engine provide integration with the ORCID (Open Researcher and Contributor ID) system.  ORCID provides a system to uniquely identify scholars across all disciplines, thus disambiguating people with similar names in the literature.  
While everybody tends to agree that there is a compelling need for author disambiguation and that ORCID is currently providing the best platform for it, accomplishing this goal is no easy feat.  For ORCID to deliver on its promise, it is necessary that the entire ecosystem of authors, publishers, and information providers work together to further its use.  

The ADS and many publishers been supporters of ORCID from its inception and have been instrumental in promoting its adoption. Currently, we are participating in this effort in two significant ways.  First, ADS is facilitating the process of claiming papers in the ORCID system via the implementation of an "ORCID mode" in the new Bumblebee interface which allows authenticated users to send their claims to the ORCID API, thus populating their bibliography list using ADS as the reference database.
Second, ADS is leveraging ORCID data in our search system by indexing ORCID identifiers provided to us in manuscripts or claimed via the user interface.  Using these two features, a scientist interested in uniquely identifying her publications can use ADS to create a bibliography and associate it with her ORCID identifier, thus making it possible for other people to discover it in ADS.

While the current initiatives by publishers and systems such as ADS can go a long way in promoting the use of ORCID, librarians can play a significant role as well.  Several observatories and research institutions currently maintain a bibliography of their staff's publications.  Some of them (such as the CfA) have started collecting ORCID identifiers for their staff, thus being able to create a mapping between bibliographic records in their collection with the ORCIDs of staff members at their institution.  These ORCID claims, curated by a trusted party, represents a source of additional data which ADS can ingest in its system to increase its ORCID coverage, without requiring any additional action on the part of the authors.  We plan to start working on the integration of ORCIDs collected by the CfA library in the fall of 2017 and to expand these efforts futher in 2018.

\subsection{Future Plans}
\label{plans}

The effort to recast the ADS technology platform began a decade ago, when it became clear that the current infrastructure was neither scalable nor maintainable, and that the development of a new system would require a significant increase in the resources available to the project.  Over these past ten years we have been able to grow the ADS team, test new technologies, gather community feedback, and plan the complex system architecture which we are now implementing.

As is often the case, the road to this new system has presented more challenges than originally anticipated, and the transition of ADS operations to the new system has been postponed as a result of this.  We are now planning to start this transition in 2018, when use of the current ADS Classic search system will be deprecated in favor of the new ADS API and the Bumblebee user interface.  We are fully aware of the impact this change will have on our community and we are therefore planning a staged migration of our user base to minimize disruption.  Initial feedback about modifying the search habits to use the new interface and search syntax will influence the timing and specifics of a broader roll-out.  

After achieving a significant migration of our users to the new system during 2018 we will focus on migrating API usage of ADS Classic, followed by access by search engines, which end up accounting for approximately half of total ADS usage.  The current plan calls for discontinuing the ADS Classic services in 2019, pending a successful transition of all of our constituencies to the new platform.  While this is a very aggressive schedule, we believe that making the transition sooner rather than later is in the interest of all parties involved: our users, our collaborators, our team of developers and curators, and our sponsors.  Our main focus at this point is therefore meeting this goal as early as possible while minimizing the impact of this change to our constituents.

There is no doubt that our transition to a new platform may cause significant disruption to both scientists and librarians.  However, the introduction of new technology and new capabilities also offers an opportunity for updating the ADS-related tools, applications, and workflows which have been developed over the years.
One such example, mentioned in section~\ref{orcid}, is the integration of ORCID claiming into the new UI and the subsequent indexing of ORCID data in the search engine, which became possible thanks for the refactoring of the system functionality as a set of separate services talking to each other.
Another example is the development of a tool based on Google Sheets which integrates with the ADS API to facilitate the management of bibliographies by librarians.  The "ADS bibliography tool," described in \cite{adsbibtool}, requires no programming experience and can be installed as an add-on from the Google Sheets interface or directly from its \href{https://github.com/adsabs/adsBibliographyTool}{GitHub repository}.

\section{Conclusions}
\label{conclusions}

We live in a time of rapid technological change, with the expectation that this change will continue to make things steadily better for all of its consumers.
This certainly includes the field of scholarly research and publishing, where use of information technology is ubiquitous from the moment of data capture at the telescope to that of publication of scientific results.  As information providers, it is our responsibility to leverage such advances and continue to improve our services to the astronomical community in order to maximize the efficiency of the research process as well as its impact on it and society at large.

At the same time the role of librarians in research institutions has been changing, and now includes activities such as digital collections curation and management, data analytics, and publishing advice.  Much of this work can benefit from the use of a system such as ADS in a number of ways: content discovery, text mining, document classification, and bibliometric analysis are all services which we make available through our new API.  Similarly, several curation activities performed by librarians can greatly benefit ADS and its user community: from article-data linking to bibliography management to ORCID enrichment.  Having your curated data indexed in ADS extends its usefulness greatly beyond the boundaries of your local institution. In this respect, sharing your data with ADS is the first step in embracing the \href{https://www.force11.org/fairprinciples}{FAIR principles}.

Over the next several years expect to see significant changes within the ADS project.  From an access point of view, the major updates described earlier will cause both discomfort and delight as users will struggle to navigate a new user interface and learn a new search syntax before realizing the benefits of the new system and its capabilities.  A significant amount of effort will go into ensuring that all relevant content is properly represented in our system, including non-conventional items such as scientific software and high-level data products.  Existing and new initiatives in the use of persistent identifiers for people, institutions and projects will make it easier to reliably identify the contributions of both individuals and collaborations.  We believe that librarians can make significant contributions in all these areas and look forward to partnering with them.  If you have suggestions or would like to get involved please let us know by contacting us directly at \href{mailto:ads@cfa.harvard.edu}{ads@cfa.harvard.edu}.

\begin{acknowledgement}
The NASA Astrophysics Data System is operated by the Smithsonian Astrophysical Observatory under NASA Cooperative
Agreement NNX16AC86A
\end{acknowledgement}

\bibliography{accomazzi}

\end{document}